# BERT-Embedding and Citation Network Analysis based Query Expansion Technique for Scholarly Search

Shah Khalid[1, 2], Shah Khusro[3], Aftab Alam[4] , and Abdul Wahid[1]

[1]Department of Computing, National University of Sciences and Technology (NUST), Pakistan
[2]School of Computer Science, Jiangsu University, China
[3]Department of Computer Science, University of Peshawar, Pakistan
[4]Division of Information and Computing Technology, College of Science and Engineering, Hamad Bin Khalifa University, Qatar Foundation, Doha, Qatar
shah.khalid@seecs.edu.pk

**Abstract.** The enormous growth of research publications has made it challenging for academic search engines to bring the most relevant papers against the given search query. Numerous solutions have been proposed over the years to improve the effectiveness of academic search, including exploiting query expansion and citation analysis. Query expansion techniques mitigate the mismatch between the language used in a query and indexed documents. However, these techniques can suffer from introducing non-relevant information while expanding the original query. Recently, contextualized model BERT to document retrieval has been quite successful in query expansion. Motivated by such issues and inspired by the success of BERT, this paper proposes a novel approach called QeBERT. QeBERT exploits BERT-based embedding and Citation Network Analysis (CNA) in query expansion for improving scholarly search. Specifically, we use the context-aware BERT-embedding and CNA for query expansion in Pseudo-Relevance Feedback (PRF) fashion. Initial experimental results on the ACL dataset show that BERT-embedding can provide a valuable augmentation to query expansion and improve search relevance when combined with CNA.

## 1    Introduction

The publication rate of 2.5 million articles per year [2]  makes it challenging to retrieve relevant publications as too many candidates may appear in response to a given search query. The techniques to cope with this issue appeared in several publication venues [9, 10, 11, 16, 20, 27]. The most prominent among these are query expansion and CNA adopted by several recent studies [11, 12, 20]. They applied both query expansion and CNA in the strategic ranking of scholarly articles. However, none of them used both context-aware embeddings using the BERT model [5] and CNA together in the process of query expansion, which could improve search relevance if employed. To bridge this gap, we propose a novel query expansion model that leverages the BERT model and CNA to select relevant chunks for query expansion in PRF fashion.

BERT is a significant breakthrough in IR [23]; Google describes BERT as the most considerable change to its search system. This research work assumes that expanding the search query using BERT (contextualized semantically related terms) from top-n initially retrieved papers and combining it with referenced papers from their citation graph can bring more relevant results. Contextualized Query Expansion for Document re-ranking has successfully improved information retrieval (IR) [7, 30]. In IR, the language used in a user query and indexed document differs in terms of verbosity, formality, and even format. Query Expansion aims to moderate this mismatch between the language used in a user query and document.

Inspired by recent developments in applying contextualized models like BERT to the document retrieval task [30], we propose QeBERT, which focuses on the deployment of BERT along with CNA for query expansion to improve scholarly search. More specifically, the proposed approach takes a few top-ranked lists of documents and its referenced document from the citation network as input for QE (e.g., from ranking model [18]) and outputs a re-ranked list based on the expanded query.

To summarize, this paper extends the existing body of knowledge regarding scholarly search by making several contributions, listed below.

- A novel approach of embedding and CNA for query expansion in academic search to improve search relevance.
- An effective inverted index and citation networks are designed and maintained from ACL papers to facilitate the proposed word embedding and CNA-based query expansion.
- A novel general architecture is proposed and materialized for word embedding and CNA-based query expansion.
- The effectiveness of the proposed approach is demonstrated on two versions of the ACL dataset using standard evaluation metrics. The proposed approach has been trained and tested on the ACL anthology dataset.

The rest of the paper is organized as: Section 2 discusses the related work, Section 3 covers a detailed discussion on the methodology; Section 4 presents the experimental setup, results & discussions are described in Section 5, and finally, Section 6 concludes the paper.

## 2 Related Work

A well-formulated search query has a significant impact on search relevance. However, a searcher is not always the expert of the field, and thus, the resulting query is often short and maybe ambiguous [3, 24]. Therefore, researchers [11, 12, 20] have developed query expansion approaches to make the initial search query enhanced and effective for bringing more relevant results. Some of them, e.g., [29], used the concept of synonymy for query expansion; others, such as [16], used external sources and other contextual details to expand queries and improve search results. Yet, others, including [9, 10], focused on making scholarly search engines more effective. Hegan et al. [10] proposed a key queries-based approach for academic search, which devises a keyword query from a few input papers of interest for retrieving related papers. For key-query, they have utilized the concept of key-queries [8]. A key query for a document returns the document in the top ranks against a (reference) search engine. However, explicitly posing a few input papers and creating a query from them is a daunting task. In our approach, we expand a query from the top papers of the initial search results and use their CNA and word embedding to add applicable terms to a user keyword query. A similar approach is adopted by the Sofia search solution [9], where a user enters a seed paper that is relevant to their research topic. The system then uses the text and the citation network to find other relevant papers. But, first, it is not fully automatic because a user needs to enter both the lower and upper bound for the citation graph of the paper. Second, all the in-cites and out-cites of the seed paper are not likely to be totally relevant to the topic of interest. Our experiment expands the query from the top relevant papers using both the citation network and word embedding.

Xiong, Power, and Callan [27] proposed explicit semantic ranking that uses embeddings based on knowledge graphs to rank research publications. They exploited the query log, Semantic Scholar's corpus, and freebase to build the knowledge graph holding concepts, descriptions, contextual relations, author-venue relations, and embeddings trained from the graph's structure. For query and document representation, learning to rank (L2R) was exploited. The published results show that the approach performed well comparatively.

The ranking is also incorporated by other means; the most prominent and widely adopted among them is the query expansion via PRF and URF (user relevance feedback) [11, 12]. In these papers [11, 12], the authors used both query expansion and CNA to improve the relevance of scholarly search. However, they were unable to consider word embedding for query expansion along with CNA. We believe, like the literature [7, 13, 20], word embedding can find semantically related terms from relevant papers.

Researchers believe that query expansion is more effective in several IR scenarios [4]. They generally perform better than their global counterparts [6, 28]. This holds for word embedding as well. The local word embedding performs well in word similarity tasks [1, 25]. In contrast, we believe that local word embedding may perform well by considering these assumptions. First, the word embeddings trained from the ACL corpus may capture the semantics of related terms better and find semantically relevant domain-specific terms compared to those trained from Wikipedia and other external resources. Second, applying the method for ranking terms in a colossal corpus may not be consistent [26]. Finally, the word embeddings trained from a domain-specific corpus can provide better performance [26]. The BERT-Base model works well comparatively in some domains, especially in QE and retrieval [30]. In the realm of scholarly search, the authors of [20] were limited to exploit the full CAN of the top PRF documents; they only consider the references and co-author network of the paper. Also, they did not exploit CNA and the BERT model in both the initial and final retrieval results. We believe that CNA and BERT-word embedding can play a vital role in the strategic ranking of the retrieval [12]. They indexed only 10000 ACL papers, we indexed 23058 papers, from which our proposed solution trains the BERT model, a different way of term ranking. The goal is to investigate the use of word embeddings using BERT and CNA for query expansion to improve the search relevance in scholarly retrieval systems.

## 3      Methodology

This paper proposes a query expansion approach for scholarly search by architect the best of the BERT-embedding model and CAN. An abstract architecture of the proposed QeBERT is shown in Figure 1. QeBERT encompasses three steps: preprocessing, BERT-Model training, and QE-based re-ranking. First, natural language processing techniques are employed on the ACL dataset to clean the textual content, which is then used to train the BERT model, and the embedding representations are learned [15]. Next, the trained model selects top $n$ similar terms from the top $n$ papers and their cited papers using CNA; the new terms are selected based on their similarity to the search query or its terms. Then, a term re-weighting might be performed to update the new query. Finally, the newly expanded query is used to retrieve a set of top-ranked 100 papers for each query using BM25 as a weighting scheme [22]. These steps are explained further in the following subsections.

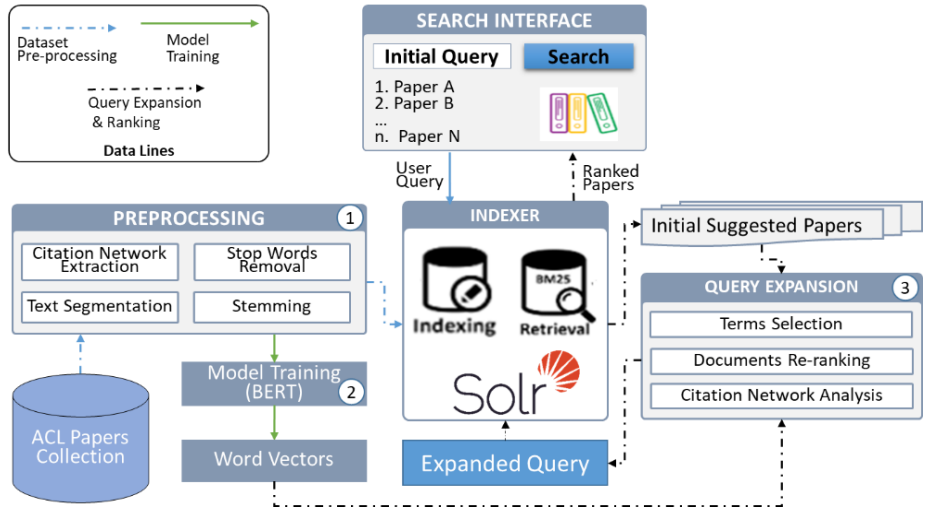

**Fig. 1.** A schematic diagram of the proposed QeBERT.

### 3.1      Pre-processing

The selected corpus is cleaned and brought to a consistent format for ease in processing. For this purpose, we used a Python-based library, known as the natural language toolkit (NLTK), in several steps. First, we extracted the citation network from the dataset by detecting and extracting the relationship between papers. The resulting citation network is a directed graph $GP = (V, E)$, where $V$ is the set of nodes representing papers $V = \{P_1, P_2, P_3, \ldots, P_n\}$ and $E$ is the set of edges $E = \{e_{ij}, 1 \leq i \leq n, 1 \leq j \leq n, \ i \neq j\}$ that shows the citation connections between papers. For instance, if a paper $P_i$ cites paper $P_j$, then there should be a directed edge $e_{ij}$ from paper $P_i$ to paper $P_j$, as shown in Figure 2 [12]. The analysis of this network is then used in query expansion, discussed in subsection 3.3, i.e., the QE is further augmented by the citation graph of the relevant PRF papers re-ranked by using trained BERT. Second, the BERT's tokenizers are used for text segmentation that split the text into sentences and sentences into words. Third, all the non-informative words are removed using the stop words list of the Apache Solr. Finally, Porter Stemmer [17] is applied for stemming.

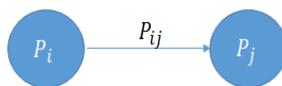

**Fig. 2.** A directed graph showing paper $P_i$ cites paper $P_j$ [12]

### 3.2    BERT-Model Training

Word embeddings refer to techniques that are used to produce vector representations of words [14]. Compared to traditional methods, the context-aware semantically similar terms are much closer and may improve search relevance. This work uses BERT for computing embeddings, where the BERT-Small variant with fewer parameters is employed to produce word vectors. We used the preprocessed ACL corpus resulting from step-1 to train the model. The model needs several parameters that can be set to improve its results. These include the number of hidden layers, the hidden embedding size, and the number of attention heads. The details about these parameters can be found in [30]. According to [30], for a small dataset, a hidden layer can be set to 4. We also set it to 4; as the ACL dataset is small, the remaining training parameters were set as: the hidden embedding size was set to 256, and the number of attention heads is also set to 4.

### 3.3    Indexing, QE-based Ranking, and Retrieval

This study used Apache Solr for indexing and retrieval, a flexible, high-performance, and scalable IR platform [6] with BM25 as the baseline weighting scheme. There are four phases in this step. First, the initial results set is obtained for a given query using BM25. Second, the initial results set is re-ranked using the BERT model described in Section 3.2. Third, the top 4 documents and their relevant referenced papers from the citation graph are considered for context-aware semantically related terms. Fourth, the final results set is obtained using selected expansion terms. The initial results obtained from BM25 are re-ranked by a fine-tuned BERT using Equation 1.

$$\text{sim (q,d)} = \text{BERT\_M}(q, d\ ) \quad (1)$$

In Equation 1, for a given query q, BERT_M assigns a similarity score 'sim' for a document d; the initial results list is then re-ranked using these scores. Now to find context-aware semantically related terms for QE, let $Q$ be the original search query having terms $q = \{q_1, q_2, \dots, q_n\}$. Let $N = \{n_1, n_2, \dots, n_n\}$ be the set of candidate terms selected from the top four retrieved (PRF) documents and their relevant referenced documents from the citation graph using the trained BERT Model from section 3.2 to score each candidate term, as indicated in Equation 2. The candidate terms are sorted in decreasing order according to their similarity scores. The top-k terms from the candidate set with the highest scores are selected. These k terms, which are the output from phase two, serve to refine the feedback information in the feedback documents from phase one.

$$\text{sim (qi, nj)} = \text{BERT}(\vec{q}_i, \vec{n}_j) \quad (2)$$

Where $q_i$ is an original query term $i$, and $n_j$ is the selected term $j$. The weight of each candidate expansion term is obtained based on its frequency and context-aware similarity with the query terms as described in Equation 3, where $f$ is the frequency of the term $n_j$.

$$\text{w ( nj)} = \text{f} \times \text{sim ( qi,nj )} \quad (3)$$

The top m terms are chosen as the final expansion terms, where $n$ is the original query terms. Moreover, a reweighting step for boosting the original query terms is applied. To this end, we chose to attribute a weight of $w = 1$ to the original query terms and a weight of $w = 0.7$ to the new expansion terms. The newly expanded queries are then used in document retrieval and ranking steps. The inclusion of the semantically related term $q_i$ in search formulation with weight $w$ may increase the effectiveness of search results (under the assumption of terms from the citation network of the top four PRF-relevant papers).

## 4 The Experimental Setup

The ACL dataset, used in the experiments of this study, holds 23,058 research articles, which are scientific publications with additional details in Table 2 [19]. The experimentation is performed on two versions of the ACL dataset. One version includes 10000 papers with 82 research questions and relevance assessments [21], as described in Table 1. The second version has 23058 documents having no relevance judgments. We carried out the manual assessment for the latter's version of the dataset, i.e., three Ph.D. students were hired to make relevance judgments for sixty search queries [11, 12]. A five-graded relevance scale was used for relevance judgments. These include 4 – highly relevant, 3 – relevant, 2 – fairly relevant, 1 – marginally relevant, and 0 – not relevant. For additional details, please read the relevant papers [11, 12]. In addition to Apache Solr for indexing and ranking using BM25, a MySQL database was used to store the citation network records along with a Perl script for providing the citation network of the top four PRF documents for candidate terms selection.

**Table 1.** Statistics about the ACL dataset with relevance Judgments

| Topics | Vocabulary Size | Indexed Docs | Average Relevant Docs |
|--------|-----------------|--------------|-----------------------|
| 82 | 32490 | 9793 | 23.67 |

Table 2. Statistics about the ACL dataset

| Item | Statistics |
|------|-----------|
| Papers | 23,058 |
| Paper Citations | 124857 |
| Venue | 350 |

For the proposed BERT-word-embedding-based-QE, the BERT model has been used to re-rank the top-100 documents retrieved in phase one in phase two. In the third phase, we set k = 4 to select the top-ranked documents from the search results of phase two, from which top 10 terms are selected for expansion, i.e., set term length m = 10 is used. In the extended version of the paper, we will examine the use of different k and m, namely, k = [10; 15; 20] and m = [10; 15; 20], studying the influences of different configurations in both Word Embedding and CNA in the effectiveness of scholarly search.

## 5    Results and Discussion

To explain the results produced by this study, three notations are used. These include baseline, representing the retrieval approach without using BERT-embedding in query expansion; BM25, the same configured method without query expansion; and BERT-QE using citation analysis, describing the proposed system. We present the analysis of both versions of the ACL dataset. Figure 3 demonstrates the results of version-1 having 10000 papers by considering user relevance judgments. Figure 4 shows the performance comparison regarding precision at $k$ (P@5, P@10, P@20) and Normalized Discounted Cumulative Gain (nDCG@10) of version-2 having about 23000 paper. Regarding precision at all three values, the proposed method performed better than the BM25 and baseline run. The same can be observed regarding nDCG@10.

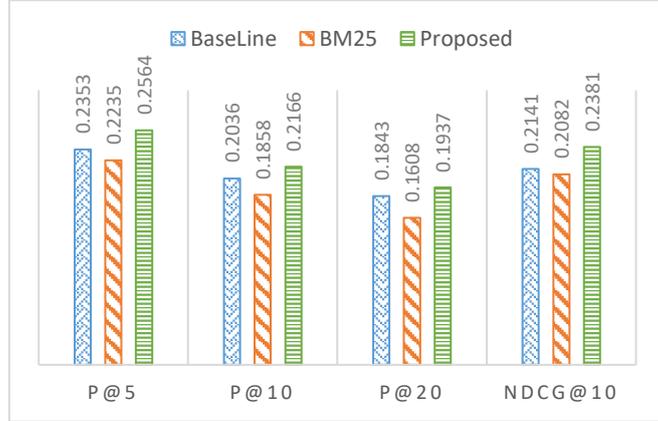

**Fig. 3.** Version-1: Results comparison of BM25, baseline, and the proposed approach

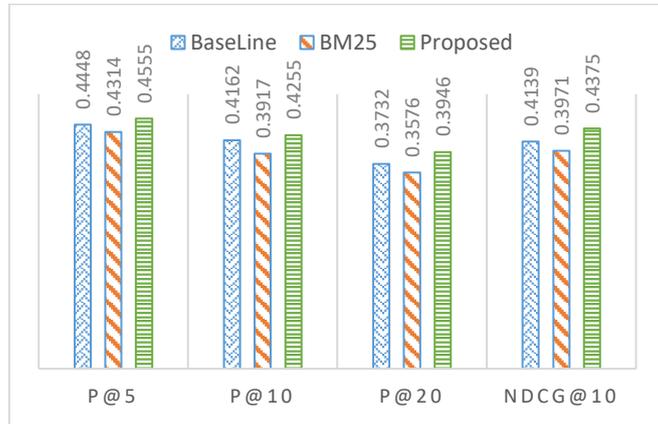

**Fig. 4.** Version-2: Results comparison of BM25, baseline, and the proposed approach

We notice that in the proposed approach, P@5 performs well then both the baseline approach and BM25. However, at P@20 (shown in Figure 3 and Figure 4), and nDCG@10 the proposed method brings slight improvements over the baseline. We hypothesize that the size of the dataset may cause this. If the entire scholarly dataset is considered, then it may utilize the trained embedding and citation network more effectively in query expansion. However, the analysis of the results indicates that considering the citation network for term selection can find more relevant candidate terms for query expansion. Figures 3 and 4 also demonstrate that as the dataset's size increases, the proposed approach's performance increases. We believe that better performance could be achieved with larger datasets.

## 6    CONCLUSION

This paper studied the impact of query expansion by BERT-word embedding and CNA on the search relevance of scholarly documents. We hypothesized that word embedding could improve search relevance. By experimenting with the ACL anthology dataset, we found improved performance against the classical IR approaches to scholarly search with BM25 as the weighting scheme and the traditional query expansion approaches that use word embedding but no CNA. From these findings, we have several takeaway lessons for future work. We plan to do extensive experimentations by considering larger datasets with BERT-Large and BERT-Base models with different configurations, i.e., k = [10; 15; 20] and m = [10; 15; 20] and so on. We also plan to try other weighting schemes and improve CNA and word embeddings for query expansion.

## References

1. Aklouche B, Bounhas I, Slimani Y (2018) Query Expansion Based on NLP and Word Embeddings. In: TREC.
2. Beel J, Gipp B, Langer S et al. (2016) paper recommender systems: a literature survey. International Journal on Digital Libraries 17:305-338
3. Clement J (2017) Statista: Average number of search terms for online search queries in the united states as of august 2017. In:
4. Dalton J, Naseri S, Dietz L et al. (2019) Local and global query expansion for hierarchical complex topics. In: European Conference on Information Retrieval. Springer, p 290-303
5. Devlin J, Chang M-W, Lee K et al. (2018) Bert: Pre-training of deep bidirectional transformers for language understanding. arXiv preprint arXiv:1810.04805
6. Diaz F, Mitra B, Craswell N (2016) Query expansion with locally-trained word embeddings. arXiv preprint arXiv:1605.07891
7. Gerth T (2021) A Comparison of Word Embedding Techniques for Similarity Analysis.
8. Gollub T, Hagen M, Michel M et al. (2013) From keywords to keyqueries: Content descriptors for the web. In: Proceedings of the 36th international ACM SIGIR conference on Research and development in information retrieval. p 981-984
9. Golshan B, Lappas T, Terzi E (2012) Sofia search: a tool for automating related-work search. In: Proceedings of the 2012 ACM SIGMOD International Conference on Management of Data. ACM, p 621-624
10. Hagen M, Beyer A, Gollub T et al. (2016) Supporting scholarly search with keyqueries. In: European Conference on Information Retrieval. Springer, p 507-520
11. Khalid S, Wu S (2020) Supporting Scholarly Search by Query Expansion and Citation Analysis. Engineering, Technology & Applied Science Research 10:6102-6108
12. Khalid S, Wu S, Alam A et al. (2021) Real-time feedback query expansion technique for supporting scholarly search using citation network analysis. Journal of Information Science 47:3-15
13. Mai G, Janowicz K, Yan B (2018) Combining Text Embedding and Knowledge Graph Embedding Techniques for Academic Search Engines. In: Semdeep/NLIWoD@ ISWC. p 77-88
14. Mikolov T, Chen K, Corrado G et al. (2013) Efficient estimation of word representations in vector space. arXiv preprint arXiv:1301.3781
15. Mikolov T, Sutskever I, Chen K et al. (2013) Distributed representations of words and phrases and their compositionality. arXiv preprint arXiv:1310.4546

16. Milliken LK, Motomarry SK, Kulkarni A (2019) ARtPM: Article Retrieval for Precision Medicine. Journal of biomedical informatics 95:103224
17. Porter MF (2006) An algorithm for suffix stripping. Program
18. Qiao Y, Xiong C, Liu Z et al. (2019) Understanding the Behaviors of BERT in Ranking. arXiv preprint arXiv:1904.07531
19. Radev DR, Muthukrishnan P, Qazvinian V et al. (2013) The ACL anthology network corpus. Language Resources and Evaluation 47:919-944
20. Rattinger12 A, Le Goff J-M, Guetl C (2018) Local word embeddings for query expansion based on co-authorship and citations.
21. Ritchie AH, Bill (2017) "Bob, the ACL Anthology test collection", Mendeley Data, V1, . In:
22. Robertson SE, Walker S (1994) Some simple effective approximations to the 2-poisson model for probabilistic weighted retrieval. In: SIGIR'94. Springer, p 232-241
23. Sabharwal N, Agrawal A (2021) Future of BERT Models. In: Hands-on Question Answering Systems with BERT. Springer, p 173-178
24. Spink A, Wolfram D, Jansen MB et al. (2001) Searching the web: The public and their queries. Journal of the American society for information science and technology 52:226-234
25. Wang Y, Huang H, Feng C (2019) Query expansion with local conceptual word embeddings in microblog retrieval. IEEE Transactions on Knowledge and Data Engineering
26. Wang Y, Liu S, Afzal N et al. (2018) A comparison of word embeddings for the biomedical natural language processing. Journal of biomedical informatics 87:12-20
27. Xiong C, Power R, Callan J (2017) Explicit semantic ranking for academic search via knowledge graph embedding. In: Proceedings of the 26th international conference on world wide web. International World Wide Web Conferences Steering Committee, p 1271-1279
28. Xu J, Croft WB (2017) Quary expansion using local and global document analysis. In: Acm sigir forum. ACM New York, NY, USA, p 168-175
29. Yue L, Fang H, Zhai C An empirical study of gene synonym query expansion in biomedical information retrieval. Information Retrieval 12:51-68
30. Zheng Z, Hui K, He B et al. (2020) BERT-QE: contextualized query expansion for document re-ranking. arXiv preprint arXiv:2009.07258